# Ground State Energies of Interacting Electrons from Projected Densities of Transitions

Roger Haydock
Department of Physics and Materials Science Institute
University of Oregon
Eugene, OR 97403-1274

For interacting electrons in solids, Heisenberg's equation is used to calculate the distribution in energy of transitions induced by adding an electron to an atomic-like spin orbital. This is the projected density of transitions which includes transitions between ground states, as well as between other states differing by one electron. The energy of a ground state is then calculated as the sum of the least energies of transitions starting with the ground state of no electrons and adding one electron with each transition. This method is applied to the construction of ground states for a simple model of the hydrogen molecule, and to interaction effects on the relative cohesion of HCP and FCC structures of transition metals.

## I. INTERACTING ELECTRONS

Calculating the electronic contribution to the cohesive energy of solids continues to challenge electronic structure theory. Density functional theory extends the independent electron approach by including interaction energies as a functional of the total density of independent electrons. However, there is still a significant gap between the energies important in many phenomena and those which can be calculated from density functional theory. The problem is that the exchange and correlation functional is non-local in position, non-linear in the density, and lacks a scheme for systematic approximation.

The projected density of transitions (PDoT) [1,2] for interacting electrons is a generalization of the projected density of states for independent electrons. It rigorously describes systems of electrons with screened (short-range) interactions, and its expansion in local environments can be made to converge rapidly with an appropriate boundary condition. In this paper, transitions are projected on an operator which creates an electron in some localized spin orbital, producing a PDoT which is a density of transitions between states of n and n+1 electrons. Among these projected transitions there are individual transitions in molecules, or bands of transitions in solids, between ground states of n and n+1 electrons. These ground state transitions can be identified as the lowest energy transitions which add one electron at a time starting with the trivial ground state of no electrons. Ground state energies and even the ground states themselves can be constructed from this sequence of ground state transitions.

The rest of the paper is organized into eight further sections with the equations defining projected transitions presented in Sec. II, followed by the relations between projected transitions, ground states, and their energies. Then there is an example using projected transitions to calculate the ground states and their energies for different numbers of electrons in a simple model of the hydrogen molecule. In Sec. IV the recursion method [3] is reviewed and applied to the PDoT for large systems. This is followed by a discussion of the local expansion of the PDoT and the boundary condition which leads to rapid convergence. In Sec. VIII there is an example of the calculation of interaction effects on the structural stability of transition metals, and the final section surveys other applications of projected transitions.

## II. PROJECTED TRANSITIONS

There is a (stationary) transition between every pair of (stationary) states of a system, however for most interesting systems there are far too many to calculate or even use if they could be calculated. The purpose of projecting transitions is to limit a calculation to just those needed for some physical quantity, and this is done by constructing only the transitions which evolve from a single (microscopic) disturbance which determines the projection. Mathematically, the disturbance is represented by an operator, and its evolution in time is generated by the Heisenberg equation with that operator as an initial condition.

A system of electrons is described by Hamiltonian **H**, which consists of the kinetic energy of the electrons, their interactions with fixed ions, their interactions with one another, and possible interactions with external fields. For simplicity take the system to be large but finite so that the states and hence transitions between states are all discrete. Now choose an initial operator **c**†. For ground state energies, a good choice is the creation operator for an electron in a particular spin orbital. Heisenberg's equation for the evolution of this operator is,

$$[\mathbf{H}, \mathbf{c}\dagger(t)] = -i\hbar\, d/dt\, \mathbf{c}\dagger(t), \tag{1}$$

where the initial condition is that **c**†(0) is **c**†. The initial operator **c**† evolves into a superposition of just a tiny fraction of all possible transitions of the system, the projected transitions for **c**†.

Taking the Fourier transform resolves **c**†(t) into stationary transitions which are the projected transitions $\{\psi_\alpha\}$ of **c**† with corresponding energies $\{\varepsilon_\alpha\}$. In terms of projection operators $\{\mathbf{P}_\zeta\}$ for a complete set of orthonormal stationary states with energies $\{E_\zeta\}$, the projected transitions may be written as,

$$\psi_\alpha = \sum_{\zeta\eta} \mathbf{P}_\zeta\, \mathbf{c}\dagger\, \mathbf{P}_\eta,\ \text{for}\ \varepsilon_\alpha = E_\zeta - E_\eta, \tag{2}$$

where the sum is over all pairs of states of the system. Note that the projected transitions are non-degenerate because all degenerate transitions $\mathbf{P}_\zeta\, \mathbf{c}\dagger\, \mathbf{P}_\eta$ contribute to a single projected transition. In very simple examples, such as in Sec. IV, the stationary states of the Hamiltonian can be calculated and Eq. 2 used to find the projected transitions. For large systems other methods are discussed in Sec. V.

The stationary transitions of the Heisenberg equation are stationary operators, analogous to the stationary states of the Schrödinger equation. The PDoT is the probability distribution of stationary transitions evolving from the initial operator in the same sense that the projected density of states is the probability distribution of stationary states evolving from some initial state. The PDoT is constructed from the projected transitions in Eq. 2 by taking the distribution in energy of the projected transitions, weighted by the sum of squared magnitudes of the matrix elements,

$$\mu(\varepsilon) = \sum_\alpha \langle \psi_\alpha\dagger\, \psi_\alpha \rangle\, \delta(\varepsilon - \varepsilon_\alpha), \tag{3}$$

where <...> indicates an inner product which in this work is the normalized trace over all states with a given electron density, then averaged over all electron densities.

The PDoT in Eq. 3 should not be confused with a thermal average, because the stationary transitions all have infinite lifetimes which they would not if the system were coupled to a thermal bath. A more helpful interpretation of the PDoT in Eq. 3 is as the probability distribution in energy for the decay of a bare electron into one dressed with electron-hole pairs.

## III. STATES AND ENERGIES

For the construction of ground states and energies, suppose the initial operator $\mathbf{c}^\dagger$ has a non-zero matrix element between every pair of ground states differing by one in their number of electrons. According to Eq. 2, this means that every ground state transition induced by adding an electron will contribute to one of the projected transitions. Whether this is true in specific cases depends on symmetry which will be discussed at the end of Sec. V.

Now suppose that $\{\psi_\alpha\}$ is the set of all projected transitions generated by $\mathbf{c}^\dagger$, and $\{\varepsilon_\alpha\}$ is the set of energies of these projected transitions. Since the $\{\psi_\alpha\}$ include contributions from states which are not ground states, they must be projected again, this time onto ground states. The simplest ground state is that of zero electrons, $\Psi_0$, the empty state. This can be used to construct the one-electron ground state $\Psi_1$ by multiplying $\Psi_0$ by the projected transitions $\{\psi_\alpha\}$. For transitions from states other than $\Psi_0$, this will give zero, and of those which are non-zero because they are from $\Psi_0$, one will have the lowest energy because the projected transitions are non-degenerate. Designate this projected transition to be $\psi_1$ with energy $\varepsilon_1$. The one electron ground state is now,

$$\Psi_1 = \psi_1 \Psi_0, \tag{4}$$

and its energy is $E_1=\varepsilon_1$. Note that because $\psi_1$ may contain only a small component of the ground state transition, the new ground state is not normalized, but that doesn't matter.

Given the n-electron ground state $\Psi_n$, with energy $E_n$, the same thing can be done to construct the n+1-electron ground state $\Psi_{n+1}$. Multiply $\Psi_n$ by the various transitions $\{\psi_\alpha\}$ and of those which give non-zero products, designate the one with the lowest energy to be $\psi_{n+1}$ with energy $\varepsilon_{n+1}$. Again, the ground state transition may be only a small component of $\psi_{n+1}$, so the new ground state,

$$\Psi_{n+1} = \psi_{n+1} \Psi_n, \tag{5}$$

is not normalized, but its energy is still,

$$E_{n+1} = E_n + \varepsilon_{n+1}. \tag{6}$$

Note that if $\psi_{n+1}$ is taken to be other than the one with the lowest energy, a stationary excited state of the system results.

By this process, an N-electron ground state and its energy are constructed, and the results summarized as a product and a sum,

$$\Psi_N = \psi_N \psi_{N-1} \ldots \psi_2 \psi_1 \Psi_0, \tag{7}$$

and

$$E_N = \varepsilon_N + \varepsilon_{N-1} + \ldots + \varepsilon_2 + \varepsilon_1. \tag{8}$$

For independent electrons, Eq. 7 becomes a determinant which is the product of creation operators for the one-electron states, and Eq. 8 becomes the sum of one-electron energies. In the interacting case the product of projected transitions is the generalization of the determinant, and the sum of transition energies is the generalization of the sum of one-electron energies.

## IV. A MOLECULAR EXAMPLE

The Hubbard model for a hydrogen molecule [4] is a familiar example of a simple interacting system which can be solved by hand and is used here to illustrate the calculation of ground states and energies for interacting systems. The model has four spin orbitals, $\varphi_{\alpha\sigma}$, where $\alpha$ is A or B depending on whether the orbital resides on one or the other of the hydrogen atoms, $\sigma$ is ↑ or ↓ depending on the spin of the orbital, and each of these can be occupied by at most one electron. The Hamiltonian has off-diagonal matrix elements $-h$ for an electron to hop between orbitals of the same spin on different atoms, and diagonal matrix elements $U$ and $2U$ for states in which both spin orbitals are occupied on one or both of the atoms, respectively.

With the exception of two states, the Hamiltonian is diagonal in determinants of bonding and anti-bonding spin orbitals. The exception is that the doubly occupied bonding orbital interacts with the doubly occupied anti-bonding orbital. In terms of the sum and difference of these two determinants, the Hamiltonian becomes a 2-by-2 matrix with diagonal elements 0 and $U$, together with off-diagonal elements both $-2h$. The ground state of this matrix has energy $\frac{1}{2}[U - (U^2 + 16 h^2)^{1/2}]$ with component $a$ of the state with zero diagonal element and $b$ for the state with diagonal element $U$, and the excited state has energy $\frac{1}{2}[U + (U^2 + 16 h^2)^{1/2}]$ with components $b$ and $-a$ respectively.

Convenient quantum numbers for the states of this system are N, the number of electrons, D, the number of bonding electrons minus the number of anti-bonding electrons (for corresponding independent electron states), S, the total electronic spin, and $S_z$, the $z$-component of electronic spin. There are 16 states, given in Table I.

From Eq. 2, the projected transitions for $c_{A\uparrow}^\dagger$ in the Hubbard hydrogen molecule are sums of degenerate transitions, with weights which are squared magnitudes of the matrix elements of $c_{A\uparrow}^\dagger$ for the transition. Note that there is an operator $c_{A\downarrow}^\dagger$ which is equivalent to $c_{A\uparrow}^\dagger$ under spin rotation. As a result, for every projected transition generated by one of these, there is a degenerate spin rotated transition generated by the other. This is a simple example of the consequences of symmetry, but it is necessary to use both sets of projected transitions in this example.

Although there are 256 independent transitions, the matrix element of $c_{A\uparrow}^\dagger$ is zero for all but 24 of them, and of those 24 there are only 8 different energies. Here projection reduces the dimension of the problem from 256 to 8. As a result, each of the projected transitions is a combination of several transitions between states. This is shown in Table II, where the transitions contributing to each of the projected transitions are listed along with their weights. Note that every projected transition is a combination of at least two and as many as four transitions between states, and that the weights vary from one transition to another. Note also that each of these projected transitions has a spin rotated partner, not listed in Table II.

Table I. Quantum numbers and energies of states of a Hubbard model for the hydrogen molecule.

| Electron Number N | Bonding Difference D | Spin S | Degeneracy g | Energy E |
|---|---|---|---|---|
| 0 | 0 | 0 | 1 | 0 |
| 1 | +1 | ½ | 2 | $-h$ |
| 1 | -1 | ½ | 2 | $h$ |
| 2 | +2 | 0 | 1 | $[U - (U^2 + 16 h^2)^{1/2}] /2$ |
| 2 | 0 | 0 | 1 | 0 |
| 2 | -2 | 0 | 1 | $[U + (U^2 + 16 h^2)^{1/2}] /2$ |
| 2 | 0 | 1 | 3 | 0 |
| 3 | +1 | ½ | 2 | $U - h$ |
| 3 | -1 | ½ | 2 | $U + h$ |
| 4 | 0 | 0 | 1 | $2U$ |

Now apply the procedure from Sec. III for constructing the ground states and energies of the model. The first step is to find the projected transition with the lowest energy giving a non-zero product on the empty state |0, 0, 0, 0>. There are two projected transitions, non-zero on the empty state. The lowest energy is $-h$ and its product with the empty state is $(1/\sqrt{2})$|1, 1, ½, ½> which is half of a Kramers doublet, the other state being |1, 1, ½, -½>.

In order to construct the 2-electron ground state, compare adding an up spin to the up spin ground state, for which the minimum energy is $h$, with adding an up spin to the down spin ground state, for which the minimum energy is $½ [U - (U^2 + 16 h^2)^{1/2}] + h$. The latter is less for all values of the parameters, so the product of the two lowest projected transition on the empty states is the un-normalized 2-electron ground state, $(a + b) /(2\sqrt{2})$ |2, 2, 0, 0>, where $a$ and $b$ are coefficients defined above. The energy of this ground state is the sum of the energies of the two projected transitions used to construct it, $½ [U - (U^2 + 16 h^2)^{1/2}]$. It is important that the projected transition, which has the lowest energy and a non-zero product with the previous ground state, gives the new ground state and its energy even though the transition between ground states is

Table II. Projected transitions for the Hubbard hydrogen molecule.

| Transition Energy $\varepsilon$ | Transitions $\|N', D', S', S_z'\rangle\langle N, D, S, S_z\|$ | Weight $\|\langle N', D', S', S_z'\|c_{A\uparrow}^\dagger\|N, D, S, S_z\rangle\|^2$ |
|---|---|---|
| $\frac{1}{2}[U - (U^2 + 16h^2)^{1/2}] - h$ | $\|2, 2, 0, 0\rangle\langle 1, -1, \frac{1}{2}, -\frac{1}{2}\|$ | $(a - b)^2/4$ |
| | $\|3, 1, \frac{1}{2}, \frac{1}{2}\rangle\langle 2, -2, 0, 0\|$ | $(a - b)^2/4$ |
| | | |
| $-h$ | $\|1, 1, \frac{1}{2}, \frac{1}{2}\rangle\langle 0, 0, 0, 0\|$ | $\frac{1}{2}$ |
| | $\|2, 0, 1, 0\rangle\langle 1, -1, \frac{1}{2}, -\frac{1}{2}\|$ | $\frac{1}{4}$ |
| | $\|2, 0, 0, 0\rangle\langle 1, -1, \frac{1}{2}, -\frac{1}{2}\|$ | $\frac{1}{4}$ |
| | $\|2, 0, 1, 1\rangle\langle 1, -1, \frac{1}{2}, \frac{1}{2}\|$ | $\frac{1}{2}$ |
| | | |
| $\frac{1}{2}[U - (U^2 + 16h^2)^{1/2}] + h$ | $\|2, 2, 0, 0\rangle\langle 1, 1, \frac{1}{2}, -\frac{1}{2}\|$ | $(a + b)^2/4$ |
| | $\|3, -1, \frac{1}{2}, \frac{1}{2}\rangle\langle 2, -2, 0, 0\|$ | $(a + b)^2/4$ |
| | | |
| $U - h$ | $\|3, 1, \frac{1}{2}, \frac{1}{2}\rangle\langle 2, 0, 1, 0\|$ | $\frac{1}{4}$ |
| | $\|3, 1, \frac{1}{2}, -\frac{1}{2}\rangle\langle 2, 0, 1, -1\|$ | $\frac{1}{2}$ |
| | $\|3, 1, \frac{1}{2}, \frac{1}{2}\rangle\langle 2, 0, 0, 0\|$ | $\frac{1}{4}$ |
| | $\|4, 0, 0, 0\rangle\langle 3, -1, \frac{1}{2}, -\frac{1}{2}\|$ | $\frac{1}{2}$ |
| | | |
| $h$ | $\|1, -1, \frac{1}{2}, \frac{1}{2}\rangle\langle 0, 0, 0, 0\|$ | $\frac{1}{2}$ |
| | $\|2, 0, 1, 1\rangle\langle 1, 1, \frac{1}{2}, \frac{1}{2}\|$ | $\frac{1}{2}$ |
| | $\|2, 0, 1, 1\rangle\langle 1, 1, \frac{1}{2}, -\frac{1}{2}\|$ | $\frac{1}{4}$ |
| | $\|2, 0, 0, 0\rangle\langle 1, 1, \frac{1}{2}, -\frac{1}{2}\|$ | $\frac{1}{4}$ |
| | | |
| $\frac{1}{2}[U + (U^2 + 16h^2)^{1/2}] - h$ | $\|2, -2, 0, 0\rangle\langle 1, -1, \frac{1}{2}, -\frac{1}{2}\|$ | $(a + b)^2/4$ |
| | $\|3, 1, \frac{1}{2}, \frac{1}{2}\rangle\langle 2, 2, 0, 0\|$ | $(a + b)^2/4$ |
| | | |
| $U + h$ | $\|3, -1, \frac{1}{2}, \frac{1}{2}\rangle\langle 2, 0, 1, 0\|$ | $\frac{1}{4}$ |
| | $\|3, -1, \frac{1}{2}, -\frac{1}{2}\rangle\langle 2, 0, 1, -1\|$ | $\frac{1}{2}$ |
| | $\|3, -1, \frac{1}{2}, \frac{1}{2}\rangle\langle 2, 0, 0, 0\|$ | $\frac{1}{4}$ |
| | $\|4, 0, 0, 0\rangle\langle 3, 1, \frac{1}{2}, -\frac{1}{2}\|$ | $\frac{1}{2}$ |
| | | |
| $\frac{1}{2}[U + (U^2 + 16h^2)^{1/2}] + h$ | $\|2, -2, 0, 0\rangle\langle 1, 1, \frac{1}{2}, -\frac{1}{2}\|$ | $(a - b)^2/4$ |
| | $\|3, -1, \frac{1}{2}, \frac{1}{2}\rangle\langle 2, 2, 0, 0\|$ | $(a - b)^2/4$ |

only one of two components of the projected transition in each case.
    There are two projected transitions which give a non-zero result when operating on the 2-electon ground state. The lowest has energy $\frac{1}{2}[U + (U^2 + 16h^2)^{1/2}] - h$, and the resulting product is $(a + b)^2/(4\sqrt{2})\|3, 1, \frac{1}{2}, \frac{1}{2}\rangle$ is again half of a Kramer's doublet. Adding the energies of the transitions to get the energy of this ground state gives $U - h$. Finally add a fourth electron to the system. Only one projected transition has a non-zero product with the spin down, 3-electron ground state. It has energy $U + h$, and gives $(a + b)^2/8\|4, 0, 0, 0\rangle$ for the ground state making

the total energy $2U$. Each time the transition between ground states is only one of several components in the projected transition and yet it gives the new ground state and its energy. Note that in many models like Hubbard models the full states such as this one can be constructed just as easily as the empty states.

For independent electrons, the ground state to ground state transition energies for adding electrons never decrease with the number of electrons in the system. If electrons repel, these transitions energies usually increase because each new electron is excluded from the states occupied by the other electrons as well as being repelled by them. Indeed, this Hubbard model for hydrogen illustrates this behavior. The first electron goes in at $-h$, the second at $½ [U - (U^2 + 16 h^2)^{½}] + h$, which is always greater than $-h$, for $U$ greater than zero. The transition energy for the third electron is $½ [U + (U^2 + 16 h^2)^{½}] - h$, which again is greater than $½ [U - (U^2 + 16 h^2)^{½}] + h$, for $U$ greater than zero. The last electron goes in at $U + h$, again greater than the previous transition energy as long as $U$ is repulsive.

For positive values of $U$ and $h$, the ordering in energy of the projected transitions is invariant with one exception. In order of increasing energy, the first and last transitions are never ground state transitions. Just inside these extreme transitions, are the two pairs of ground state transitions, numbers 2 and 3 together with numbers 6 and 7. Numbers 4 and 5 are not ground state transitions and are the only pair which can interchange their energy ordering depending of whether $U$ is greater or less than $2h$.

The ordering of the transition energies in this example hints at the formation of bands in the projected transitions for extended systems. In the case of a molecular solid, transitions 2 and 3 might be interpreted as the bottom and top of a bonding band with a gap containing 4 and 5, and then an anti-bonding band between 6 and 7. Transitions which are not ground state transitions, numbers 1, 4, 5, and 8, lie outside the bands.

It is very important that the ground state projected transitions can contain contributions from other degenerate transitions and still produce the next ground state when multiplying the previous one. The only effect of these extra contributions on the product is to change the normalization, which is arbitrary anyway. These projected transitions add one electron to the system, because the number of electrons in a state is independent of its normalization.

## V. DIRECT CALCULATION OF PROJECTED TRANSITIONS

In the previous section all stationary transitions were calculated before Eq. 2 was used to obtain the projected transitions. If the stationary transitions were known, the problem would already be solved, so the purpose of this section is to show how the PDoT can be calculated directly from the Hamiltonian without intermediate steps.

The evolution of the projecting operator $\mathbf{c}\dagger$ is given in Eq. 1 by the time-dependent Heisenberg equation. The formal solution of this equation is,

$$\mathbf{c}\dagger(t) = exp\{i\,L\,t/\,\hbar\}\,\mathbf{c}\dagger, \qquad (9)$$

where $L$ is a super-operator defined on an operator $\mathbf{x}$ by the commutator,

$$L\,\mathbf{x} = [\mathbf{H},\,\mathbf{x}]. \qquad (10)$$

This solution only makes sense if the expansion of Eq. 9 in powers of $t$ converges at least near $t=0$, and this leads to the restriction that the Hamiltonian must be short-ranged [1]. For electrostatics this means using the screened interaction as in, for example, the Hubbard model. So, assume in what follows that **H** contains only the short ranged parts of interactions and that the series expansion of Eq. 9 converges at least for $t$ near 0.

The power series for Eq. 9 expresses $\mathbf{c}\dagger(t)$ as a linear combination of the operators $\{L^n\mathbf{c}\dagger\}$ for n = 0, 1, 2, ... . This is exactly the set of operators which spans the projected transitions for $\mathbf{c}\dagger$, or in other words, these operators span the smallest invariant subspace of operators containing $\mathbf{c}\dagger$. For the Hubbard model and other models using localized or tight-binding electronic bases, $L^n\mathbf{c}\dagger$ can be expanded as a sum in paths starting with $\mathbf{c}\dagger$ and consisting of n hops on a lattice of products of $\mathbf{c}_\alpha\dagger$ and $\mathbf{c}_\beta$ for various orbitals $\alpha$ and $\beta$. The contribution of a path to $L^n\mathbf{c}\dagger$ is the last operator on the path multiplied by the product of matrix elements of **L** between the operators along the path. This follows from matrix multiplication of the representation of **L** in the basis of products of orbital operators. The expansion in paths is a discrete version of the construction of propagators by path integrals, and in the non-interacting case $L^n\mathbf{c}\dagger$ is just a sum of paths on the lattice of electronic orbitals, see Ref. 3 for examples of this.

The $\{L^n\mathbf{c}\dagger\}$ are ill-conditioned for numerical applications due to their lack of orthogonality. However, an orthonormal basis of operators can be constructed from the $\{L^n\mathbf{c}\dagger\}$ using the average of normalized traces introduced at the end of Sec. II. The inner product between operators **x** and **y** is just $<\mathbf{x}\dagger\mathbf{y}>$ and a convenient way to construct the basis is by a Gram-Schmidt (GS) process. Because **L** is Hermitian in this inner produce, the GS is symmetric and the resulting matrix for **L** is tridiagonal. Applying the symmetric GS to $\{L^n\mathbf{c}\dagger\}$ generates the sequence of orthogonal operators $\{\mathbf{u}_0, \mathbf{u}_1, \mathbf{u}_2, ..., \mathbf{u}_n, \mathbf{u}_{n+1}, ...\}$, with the initial conditions that $\mathbf{u}_0 = \mathbf{c}\dagger/<\mathbf{c}\,\mathbf{c}\dagger>^{1/2}$, and $\mathbf{u}_{-1} = \mathbf{0}$. Subsequent operators are constructed recursively according to the following relation:

$$\mathbf{u}_{n+1} = (L\,\mathbf{u}_n - a_n\,\mathbf{u}_n - b_n\,\mathbf{u}_{n-1})/\,b_{n+1}, \tag{11}$$

where,

$$a_n = <\mathbf{u}_n\dagger\,L\,\mathbf{u}_n>,$$

$$b_{n+1} = <(L\,\mathbf{u}_n - a_n\,\mathbf{u}_n - b_n\,\mathbf{u}_{n-1})\dagger\,(L\,\mathbf{u}_n - a_n\,\mathbf{u}_n - b_n\,\mathbf{u}_{n-1})>^{1/2}.$$

Note that because of Hermiticity, $L\mathbf{u}_n$ contains only components of $\mathbf{u}_{n-1}$, $\mathbf{u}_n$, and $\mathbf{u}_{n+1}$. The process formally terminates when $b_{N+1}$ is zero, meaning that the space spanned by $\mathbf{u}_0, \mathbf{u}_1, \mathbf{u}_2, ..., \mathbf{u}_N$ is invariant under **L**. In practice the process rarely terminates, usually because the invariant subspace it too large for a complete basis to be computed, or sometimes because rounding error destroys the orthogonality of the $\{\mathbf{u}_n\}$.

The real scalars $\{a_0, a_1, a_2, ... a_n, a_{n+1}, ...\}$ and the positive real scalars $\{b_1, b_2, b_3, ... b_n, b_{n+1}, ...\}$ in Eq.11 are respectively the diagonal elements and main sub-diagonal matrix elements of a symmetric tri-diagonal matrix **J** which is a matrix representation of **L** in the basis $\{\mathbf{u}_n\}$. Because the matrix is tri-diagonal, the projected transitions can be written as,

$$\psi_\alpha = \sum_n \psi_n(\varepsilon_\alpha) \mathbf{u}_n, \qquad (12)$$

where the sum is over the basis $\{\mathbf{u}_n\}$, and the coefficients $\{\psi_n(\varepsilon)\}$ satisfy the three term recurrence,

$$\psi_{n+1}(\varepsilon) = ((\varepsilon - a_n) \psi_n(\varepsilon) - b_n \psi_{n-1}(\varepsilon))/ b_{n+1}. \qquad (13)$$

This has two linearly independent solutions which may be combined to satisfy an initial condition involving $\psi_{-1}(\varepsilon)$, or a boundary condition involving $\psi_{N+1}(\varepsilon)$ where $\mathbf{u}_N$ is the last basis element computed, or both.

The resolvent matrix, $(\varepsilon I - J)^{-1}$ (where $I$ is the identity), is simple to calculate for $J$ tri-diagonal, and the $\mathbf{u}_0$-$\mathbf{u}_0$ element of this resolvent can be expressed in terms of bounded solutions $\{\psi_n(\varepsilon)\}$ to the recurrence in Eq. 13, which can be expanded as the continued fraction,

$$R(\varepsilon) = 1/\,\varepsilon - a_0 - b_1 \psi_1(\varepsilon)/\psi_0(\varepsilon) = 1/\,\varepsilon - a_0 - b_1^2/\,\varepsilon - a_1 - b_2^2/\,\varepsilon - a_2 - \ldots - b_n^2/\,\varepsilon - a_n - \ldots. \qquad (14)$$

The imaginary part of the resolvent $R(\varepsilon)$ is negative in the upper half of the complex $\varepsilon$-plane and positive in the lower half with the only singularities in $R(\varepsilon)$ occurring on the real $\varepsilon$-line. Taking the singular part of $R(\varepsilon)$ to be directed from the lower to upper halves of the $\varepsilon$-plane, the PDoT,

$$\mu(\varepsilon) = - <\mathbf{c}\,\mathbf{c}\dagger> Sing\{R(\varepsilon)\}/ 2\pi i, \qquad (15)$$

or in other words, the PDoT is the residue of $R(\varepsilon)$ which contributes to integrals enclosing portions of the real $\varepsilon$-line, normalized to the projecting operator.

The above transformation of Heisenberg's equation, with the initial condition $\mathbf{c}\dagger$, to at tri-diagonal matrix can be interpreted as projecting the original system onto a one-dimensional, semi-infinite, tight-binding model with nearest neighbor interactions, a chain model. This reflects the one-dimensional nature of the time evolution of $\mathbf{c}\dagger$ which only spans the full space of transitions if $\mathbf{c}\dagger$ couples to every transition (no transitions are degenerate).

If $\mathbf{H}$ has symmetries, then so does $L$, and different initial operators $\mathbf{c}\dagger$ are equivalent under symmetry operations. If states are to be constructed, then projected transitions will be needed for each in-equivalent initial operator, with the projected transitions for equivalent initial operators constructed by symmetry transformations, as in Sec. IV. The density of transitions per unit volume can be calculated from a single initial operator which is the sum, within a unit volume, of in-equivalent initial operators, each weighted by the root of the number of equivalent operators in the unit volume.

The preceding is a summary of the recursion method [3], one of a large family of methods which apply various kinds of moments to a wide range of calculations. In the case of the recursion method the underlying moments are the power moments $<\mathbf{c}\, L^n\, \mathbf{c}\dagger>$. As with other moment methods and related high temperature expansions, computations are dominated by the growth of the number of terms contributing to successive moments. In principle this can be as great as n! [1], but seems to be only exponential in cases of interest. However for even simple models, this exponential growth sets an upper bound on computations of about 30 recursions, equivalent to 60 power moments.

While the calculation of 60 moments for one of these methods is an impressive achievement, it does not, on its own, give sufficient resolution to answer many important questions. What is needed is an approximation for the high order moments not calculated, which in terms of this work, are the rest of the matrix elements of $J$, the rest of the recurrence relation in Eq. 13, or the tail of the continued fraction in Eq. 14. This requires information about the nature of the system far from where $c^\dagger$ creates the electron, and is the topic discussed below.

## VI. THE LOCAL EXPANSION AND BOUNDARY CONDITION

An advantage of the recursion method is that $\{\mathbf{u}_n\}$ is an expansion of the PDoT in terms of local environments of $c^\dagger$. The first basis element $\mathbf{u}_0$ is just $c^\dagger$ itself, and its matrix element, $a_0$, is the energy of the spin orbital in which $c^\dagger$ places an electron. This is the most important parameter in the PDoT. The next basis operator $\mathbf{u}_1$ is a combination of operators which create electrons and holes in orbitals near the original one because the Hamiltonian is short ranged, and this operator determines the parameters $b_1$ and $a_1$, which are respectively the coupling of $\mathbf{u}_0$ to $\mathbf{u}_1$ and the energy of $\mathbf{u}_1$. These two parameters are the next most important in determining the shape of the PDoT.

As the recursion continues each subsequent $\mathbf{u}_n$ creates electrons and holes in orbitals which are further away from $\mathbf{u}_0$ in the precise sense that one more application of $L$ is necessary to access these operators from $\mathbf{u}_0$. For non-interacting electrons this is very similar to distance in the usual sense, but for interacting systems this distance is more abstract including the creation of electrons-hole pairs as well as spatial hops. The parameters $b_n$ and $a_n$ associated with $\mathbf{u}_n$ have less influence on the shape of the PDoT because they are further from $\mathbf{u}_0$ in this sense of locality.

The convergence of this local expansion is made precise by a theorem of von Laue [5]. It states that inside a cavity, the local density of electromagnetic modes, the distribution of intensities near a given frequency and position, depends on the shape and nature of the cavity wall in a way which decreases exponentially with the number of wave lengths to the surface of the cavity. Friedel [6] pointed out that this theorem applies to any linear wave equation, such as the Schrödinger equation where the modes are stationary states and the cavity is a finite volume of material. In this work the cavity remains a finite volume of material, and the Heisenberg equation is the linear wave equation, waves of operators or transitions, but still waves.

For an ideal cavity, the distribution of modes at the surface is the same as inside the cavity, and this of course gives the most rapid convergence with distance from the surface. The best convergence of the PDoT is obtained when the cavity is as close to ideal as possible, meaning that the material outside the cavity is as similar as possible to the material inside the cavity, or at least near its surface. This can be expressed as a kind of self-consistency, that the material and hence the PDoT changes as little as possible at the boundary.

Von Laue's theorem also applies to the calculation of a finite part of the recurrence in Eq. 13, because it is also the time-independent form of a linear wave equation, discrete rather than continuous. The conclusion is that PDoT converges exponentially with the number of wave lengths along the chain between the first basis element $\mathbf{u}_0$, which is the center of the semi-one-dimensional cavity, and the last one $\mathbf{u}_N$, which is the surface of the cavity. The basis elements beyond $\mathbf{u}_N$ describe the material outside this abstract cavity, and so for a nearly ideal cavity, the recurrence inside and outside are nearly the same.

If the PDoT consists of a single band of transitions whose width and center are known, then the continuation of the recurrence with constant matrix elements, $a_n=a$ and $b_n=b$ for n>N does well for $a$ and $b$ fit to the known band. When these matrix elements are used to extend the continued fraction, they are called the constant terminator and this is used to calculate the PDoT in Sec. VIII. However, when there are multiple bands as is usually the case for strongly interacting systems, such extensions of the recurrence are very difficult to construct.

For multiple bands, the problem of matching the interior and exterior of the effective cavity is best reformulated in terms of a boundary condition at the surface of the cavity. If the interior and exterior of the cavity match, then the boundary should not reflect any part of a wave impinging on it. The condition of no reflections can be expressed more generally as a boundary condition which maximally breaks time-reversal symmetry (MBTS) [7,8] in the sense that the two waves which satisfy the boundary condition form a time-reversal doublet with maximum current in each direction. Since there are two solutions, the boundary condition is a homogeneous quadratic in the $\{\psi_n(\varepsilon)\}$ of Eq. 13, a simple self-consistency condition.

What all this says is that the most important information about the PDoT is in its local environment with nearer environments more important than distant ones, where near means fewer powers of *L*. Provided the system is reasonably homogeneous on the scale of the 30th or so local environment, the MBTS boundary condition gives useful accuracy, even for complicated systems.

## VII. INTERACTION EFFECTS IN METALLIC COHESION

Metals have continuous electronic bands rather than discrete atomic or molecular levels, and this is reflected in the PDoT for a metal. Because the electronic states of a metal form bands, so do the transitions between them, and in the PDoT for **c**† these transitions are weighted by the probability they are induced by adding the electron. Consider a simple version of the Hubbard model for interacting electrons in an alkali metal with a cubic crystal structure, and take the initial operator **c**† to create an electron in the spin-up valance orbital of one atom. As in Sec. IV, each atom has two spin-orbitals, and the Hamiltonian consists of hopping terms with matrix element -*h* between orbitals of the same spin on nearest neighbor atoms, as well as interaction terms which add a diagonal energy *U* for each atom with both spin orbitals occupied in the state.

When *U* is zero, the PDoT for **c**† has a band which extends in energy from -6*h*, which is the lowest energy at which an electron can be added to the system, to 6*h*, the largest energy at which an electron can be added. For *U* greater than zero, the smallest energy at which the first electron can be added remains -6*h*, but now, if the system is in the ground state with just one hole, the last electron can only be added with energy 6*h*+*U* due to repulsion from the electrons already there. These limits define the band of transitions between ground states which differ by one electron. Provided *U* is small enough not to produce gaps in the band of ground state transitions, there is a ground state transition at every energy from -6*h* to 6*h*+*U*; and, although the projected transitions in this band of energies may contain components of transitions not between ground states, the ones between ground states have to be present because they have non-zero projections on **c**†.

Having established that there is a band of ground state transitions in the PDoT for this system, the arguments of Sec. III can be applied in a continuum form to determine the ground state energy. In this simple model every spin orbital is equivalent, so the PDoT for one orbital is proportional to the number of ground state transitions per spin orbital, per unit of energy. (When there are non-equivalent orbitals this density of ground state transitions is a combination of the PDoTs for these orbitals.) Each orbital holds a total of one electron, so, for a Fermi level $\varepsilon_F$, the number of electrons per spin orbital is,

$$n(\varepsilon_F) = \int_{occ} \mu(\varepsilon) \, d\varepsilon \, / \int_{band} \mu(\varepsilon) \, d\varepsilon, \tag{16}$$

where the first integral is over the occupied part of the band, from $-6h$ to $\varepsilon_F$, and the second is over the whole band from $-6h$ to $6h+U$. The transition at $\varepsilon_F$ adds one electron to the ground state with $n(\varepsilon_F)$ electrons per orbital, producing the new ground state. So, the electronic energy per spin orbital of the ground state with $n(\varepsilon_F)$ electrons per orbital is the sum of these transition energies,

$$E(\varepsilon_F) = \int_{occ} \varepsilon \, \mu(\varepsilon) \, d\varepsilon \, / \int_{band} \mu(\varepsilon) \, d\varepsilon, \tag{17}$$

where the integrals are over the same intervals as in Eq. 16. Construction of the ground state is left for future work.

Interactions make other changes in the PDoT beside the shift in the top of the band of ground state transitions. A non-zero value of $U$, even though it is small, adds terms to the commutators $\boldsymbol{L}\mathbf{u}_n$. For n=0, the interaction adds an electron-hole pair. For subsequent n, the interaction adds an electron-hole pair for every site which is occupied by a single electron or hole. This adds up to an exponential increase with n in the number of terms in $\mathbf{u}_n$. Even though the coefficients of these terms decrease as powers of $U/h$, the increase in the number of terms wins and the matrix elements of $J$ in Eq. 11 increase with n.

The consequence of increasing matrix elements in $J$ is that the PDoT develops infinite tails above and below the band of ground state transitions. This is the mechanism by which non-ground state transitions get mixed with ground state transitions in the projected transitions. The result of this is that as $U$ increases, spectral weight shifts from the band to the tails, necessitating the normalization of the band of ground state transitions in Eqs. 16 and 17 to one electron per orbital.

As in the molecular case, the PDoT must contain transitions between states other than ground states. For a start, the addition of the first electron at an energy greater than $-6h$ produces a stationary excited state. Subsequent electrons can be added while maintaining stationary excited states, however for excited states of the system the last electron must be added at an energy below $6h+U$, because that is the energy of the last ground state transition in the band. As a result, the transitions between excited stationary states are also components of the projected transitions in the band and could be constructed by adding electron at energies other than ground state energies.

For $U/h$ small, the band of ground state transitions has upper and lower edges as well as two internal van Hove singularities at the energies where the Fermi surface touches Brillouin zone boundaries. As $U/h$ increases the bottom band edge remains fixed, the top edge moves to higher energies, and other singularities including gaps can develop in the band. Singularities in the PDoT occur at energies for which the transitions change in some qualitative way. Weak singularities, like kinks in the band, correspond to mild changes in the transitions such as occur when the states encounter zone boundaries, and stronger singularities such as the formation of a gap in the band correspond to ordering in the states underlying the transitions. This can become complicated, but the principle from Sec. II remains valid, that the next ground state transition is the one lowest in energy whose product with the previous one is non-zero.

## VIII. Interaction Effects on the Structural Stability of Transition Metals

Using the results from the previous section, the effects of interactions can be included in the classic problem of comparing energies for different stacking sequences in transition metal crystal structures [9]. The paradigm of this is the comparison of hexagonally close packed (HCP) and face centered cubic (FCC) structures which are distinguished by the two possible relations between the first and third planes of close packed atoms. HCP has the third plane directly above the first and FCC does not. For these and more generals stacking structures, the atoms occupy the same volumes, with the same first neighbors, differing only in the bond angles from first to second neighbors. The d-electrons are sensitive to these subtle differences in geometry and the relative stability of a great many transition metals and their compounds seems to be explained by just the d-contribution to the cohesive energy.

There is some evidence that the repulsion between d-electrons makes a significant contribution in compounds of early and late transition metals [10], so it is worth exploring this effect in a simple model. The issue addressed here is how interactions change the relative stability of HCP and FCC for pure transition metals. The model is again a Hubbard model in which there are five d-orbitals on each atom with hopping matrix elements between orbitals on neighboring atoms expressed in terms of Slater-Koster parameters[11]. A Hubbard term with energy $U$ is added for each orbital, increasing the energies of states by this amount for each doubly occupied spatial orbital. The initial operator for this calculation is constructed so that the resulting PDoT is the total density of transitions per orbital, see the end of Sec. V.

Using a single parameter $B$, in terms of which the Slater-Koster parameter dd$\sigma$ is -2$B$, dd$\pi$ is $B$, and dd$\delta$ is zero, Ducastelle and Cyrot-Lackmann [10] calculated the first four power moments of the total density of states per orbital for non-interacting d-electrons in these structures. In this model the only difference between the two structures is that the fourth moment of the d-band is smaller in HCP. These moments can be used to evaluate the matrix elements of $J$ in Eq. 11, which in turn give the continued fraction parameters in Eq. 14. These are shown in Table III. When the interaction terms are included, the matrix elements of $J$ and the continued fraction parameters depend on $U$, as do the band widths for the two structures, shown in the last column of Table III.

The next step is to use the parameters in Table III to construct a continued fraction whose singular part gives an approximate PDoT for these structures. There are explicit parameters for the first two levels of the continued fraction in Eq. 14. The rest of the continued fraction is approximated by the condition of no reflection [7].

The PDoT obtained by Eq. 15 from the continued fractions is integrated according to Eqs. 16 and 17 to give $n(\varepsilon_F)$ and $E(\varepsilon_F)$ for each of the two structures. The difference between the electronic energy of FCC and HCP is plotted in the Fig. as a function of the band filling for $U=0$, $B$, and $2B$. Note that the differences in cohesive energies are small and that they favor FCC at the low and high extremes of band filling. This is because $b_2(U)^2$ for FCC is always larger than for HCP, so the band of transitions is wider for FCC, and the first few electrons for the empty band or the first few holes for the full band, go in at lower energies for FCC than for HCP.

As can be seen from the Fig., the effect of interactions is to reduce the relative difference between $b_2(U)^2$ for FCC and HCP. This has the somewhat subtle effect of shifting weight in the PDoT toward the band edges which then increases the ranges of low and high band filling for which the FCC structure is stable. Further increasing the interactions reduces the importance of the difference in second neighbor positions and so the cohesive energy differences become smaller.

Table III.  Tri-diagonal matrix elements in units of $B$ (see text).

|            | HCP ($U=0$) | FCC ($U=0$) | Interacting Model ($U>0$) |
|---|---|---|---|
| $a_0(U)$   | 0.0    | 0.0    | $U/3$ |
| $b_1(U)^2$ | 3.6    | 3.6    | $b_1(0)^2 + 2\,U^2/9$ |
| $a_1(U)$   | -0.361 | -0.361 | $U/3 + (a_1(0)\,b_1(U)^2 + 2\,U^3/27)/\,b_1(U)^2$ |
| $b_2(U)^2$ | 2.936  | 3.158  | $\{b_1(0)^2\,[a_1(0)^2 + b_2(0)^2 + (10/9)U^2]$ $+ (2/81)U^4 - a_1(U)^2\,b_1(U)^2\}/\,b_1(U)^2$ |

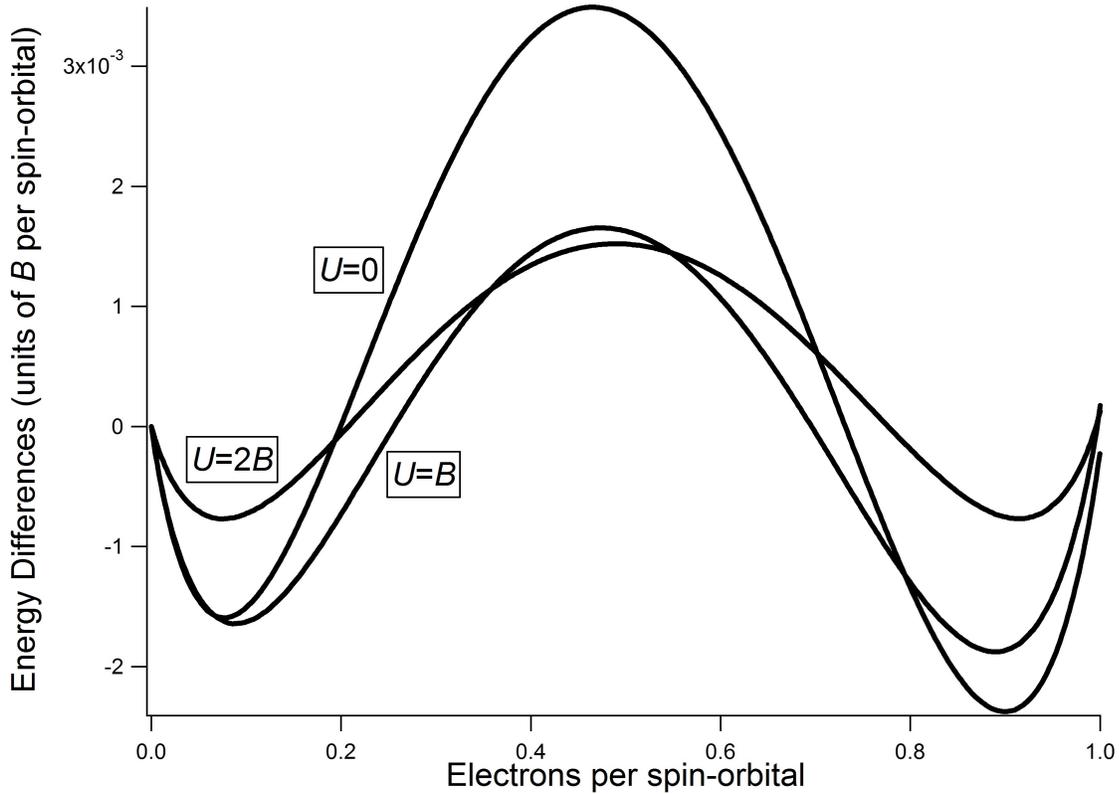

Figure.  Differences between d-band cohesive energies for FCC and HCP structures in a Hubbard model.

## IX. OTHER PROJECTIONS

If the dimension of the space of states is N, then the space of operators has dimension $N^2$, any one of which could be used to project transitions. Electron operators for atomic orbitals can be combined in two ways: They can be added; for example in a crystal a Bloch sum of such operators creates excitations with a specific crystal momentum, and the resulting PDoT is related to the band structure. Or they can be multiplied; for example spin flips produce magnons, electron-hole pairs produce plasmons, electron pairs produce Cooperons, and so forth. With all these possibilities there are many PDoTs to study.

In previous work [12] on Heisenberg models, a super-operator was constructed which added a site to the model, however for spin-½ Heisenberg models this is equivalent to adding an electron to an empty site. Unlike metallic systems, Heisenberg models are insulators, so the occupations of sites are fixed. The electron on one site can only exchange spins with its neighbors. As a result the projected transitions are much simpler than for metals.

The PDoT for the Heisenberg model consists of multiple bands extending over all energies, however, as in the metallic case, there is a special ground state which singles out one of these bands. It is a ferromagnetic ground state which can be constructed as the product of operators creating electrons with the same spin on each site. The transition energy for adding the electron to this state is just the sum of its interactions with neighboring sites. The other transitions in this band are between stationary (thermal equilibrium) states for which the added spin binds with less energy, and the top of this band is the transition for which the extra spin binds with the least possible energy, the anti-ferromagnetic ground state. The transitions in this band may be interpreted as the binding of the extra spin at various temperatures, both positive and negative.

Just as for metallic cohesion, this band of the PDoT may have internal singularities at binding energies where the projected transitions change qualitatively. If the transitions change qualitatively, so must the underlying states, and so these internal singularities must include the binding energies at which there are phase changes.

In summary, the PDoT is a compact, computable, distribution, rigorously characterizing large systems of linear equations, which in this work is the Heisenberg equation for operators acting on interacting electrons. The choice of initial operator determines which invariant subspace of transitions contributes to the PDoT, and this in combination with special states provides the key to extracting physical information. This approach applies to any theory with linear equations of motion, from electromagnetism, to quantum mechanics, even to classical mechanics formulated in terms of distributions of particles [13].

## ACKNOWLEDGEMENTS

The Author gratefully acknowledges support for this work from the Richmond F. Snyder Fund and gifts to the University of Oregon in memory of Richmond F. Snyder, as well helpful discussions with Chris Nex and James Annett.